\documentclass[12pt,preprint]{aastex}

\newcommand{\xmm}{{\em XMM-Newton}}
\newcommand{\epn}{{EPIC PN}}
\newcommand{\src}{RE~J1034+396}
\newcommand{\msun}{M\ensuremath{_\odot}}	
\newcommand{\redc}{\ensuremath{\chi^2/dof}}	

\shorttitle{Spectral variability in \src}
\shortauthors{Maitra \& Miller}
\begin{document}
\title{Evidence of a Warm Absorber that Varies with QPO Phase in the AGN \src}

\author{Dipankar Maitra\altaffilmark{1} and Jon M. Miller\altaffilmark{1}}
\affil{Department of Astronomy, University of Michigan,
    Ann Arbor, MI 48109}
\email{dmaitra,jonmm@umich.edu}

\begin{abstract}  
A recent observation of the nearby (z=0.042) narrow-line Seyfert 1 galaxy
\src\ on 2007 May 31 showed strong quasi-periodic oscillations (QPOs) in
the 0.3--10 keV X-ray flux. We present phase-resolved spectroscopy of this
observation, using data obtained by the EPIC PN detector onboard \xmm.
The ``low'' phase spectrum, associated with the troughs in the light
curve, shows (at $>4\sigma$ confidence level) an absorption edge at
$0.86\pm0.05$ keV with an absorption depth of $ 0.3\pm0.1$.  Ionized oxygen
edges are hallmarks of X-ray warm absorbers in Seyfert active galactic
nuclei (AGN); the observed edge is consistent with H-like O VIII and
implies a column density of $N_{\rm OVIII}\sim3\times10^{18}$ cm$^{-2}$.
The edge is not seen in the ``high'' phase spectrum associated with the
crests in the light curve, suggesting the presence of a warm absorber in
the immediate vicinity of the supermassive black hole which periodically
obscures the continuum emission.  If the QPO arises due to Keplerian
orbital motion around the central black hole, the periodic appearance
of the O VIII edge would imply a radius of $\sim9.4(M/[4\times10^6
\msun])^{-2/3}(P/[1 hr])^{2/3}\;r_{\rm g}$ for the size of the warm
absorber.
\end{abstract}


\section{Introduction}
\src\ is a narrow line Seyfert 1 galaxy (NLS1; see \citealt{op1985}
for definition of NLS1s). It is well known for its high soft X-ray
excess compared not only to other AGNs but also among the NLS1s
\citep{pu1995,boller1996,middleton2007}.  While the origin of this soft
excess is still not clear, the similarity of its soft X-ray spectrum with
that of stellar black hole binaries in their `high state' has been used
to postulate that \src\ harbors a comparatively low mass black hole which
is currently in a state of high mass accretion rate \citep{pounds1995}.

A recent X-ray observation of \src\ using the \xmm\ satellite made
between 2007 May 31 and 2007 June 1 showed a strong signature of
quasi-periodic variability in the 0.3--10 keV light curve \citep{g08}.
The oscillations were transient in nature since they have never been seen
in any prior observations of this source, nor did we see strong QPO-like
variability in the 0.3--10 keV light curve obtained during a subsequent
\xmm\ observation made on 2009 May 31.  It was noted by \citet{g08} that
the frequency of the observed oscillations was similar to what would be
expected if the frequencies of the quasi-periodic oscillations often
seen in Galactic black hole binaries \citep[see e.g.][for a review on
observations of QPOs in compact stellar X-ray binary systems]{vdk2006}
were to be scaled by the same factor as the mass ratio between the mass
of the supermassive black hole in \src\ and the mass of typical stellar
black holes.  Recently \citet{MiddletonDone2009} have performed detailed
comparison of the timing properties of this \xmm\ observation with that
of stellar mass black hole binaries and concluded that the observed QPO
in \src\ is similar to the 67 Hz QPO seen in GRS~1915+105 (a black hole
binary system which, like \src, also boasts a high L/L$_{\rm Edd}$).
This scaling of oscillation frequencies with mass would support
the hypothesis that accretion physics scales with mass and that the
supermassive black holes at the center of galaxies are scaled-up versions
of stellar-mass black holes.

The origin of QPOs in stellar black hole (and neutron star) binaries is
not yet well understood.  Various models have been proposed, but none can
fully explain the wealth of observed phenomenology.  QPO frequencies in
stellar black hole binaries range from Hertz to kHz, and it is extremely
difficult to carry out phase-resolved spectroscopy in these systems.
In the sparse cases \citep[e.g. in the source GRS~1915+105, reported
by][]{Morganetal1997} where it is possible to ``see'' the QPO in the light
curves, the phase of the oscillations appear to perform a random walk.
\citet{MillerHoman2005} showed that for GRS~1915+105 the flux in the Fe
K$\alpha$ line (created by hard X-ray photons irradiating a relatively
cold accretion disk) varies with the QPO oscillation phase of the 1 and
2 Hz QPOs, thus suggesting that the 1 Hz and 2 Hz QPOs may be linked to
variable reflection.  The strong variability amplitude and comparatively
large period of the oscillations in \src\ make it an ideal source to
carry out phase-resolved spectroscopy and probe changes in the spectral
energy distribution between different phases.

The origin of the soft excess in \src\ is not clear, and its time-averaged
spectral energy distribution can be well modeled by a wide variety
of models including reflection from a partially ionized accretion disc
\citep{crummy2006}, Comptonized disc emission from a low temperature disc,
ionized partial covering, or a smeared disc wind seen in absorption
\citep[see e.g.][]{middletonetal2009}.  Phase-resolved spectra could
give important insights for breaking this degeneracy.  Therefore one of
our major goals in this work is to search for signatures of variable
reflection (in the continuum and reflection lines) and/or variable
absorption features (e.g.  changes in the properties of O VII, O VIII
edges which are the hallmark of the presence of warm absorber in active
galactic nuclei).

\citet{middletonetal2009} analyzed the energy dependence of the
observed variability in the EPIC data and showed that the variability
primarily originated in the high-energy photons.  They also presented
a spectral decomposition of the EPIC MOS spectrum, averaged over
the entire observation.  Here we analyze the EPIC PN data taken
during this observation to test if there is any signature of spectral
variations between the different phases.  Due to limitations in photon
statistics (largely caused by the high pileup on the CCD as described
in \S\ref{prepare_data}; also see \citealt{middletonetal2009}) we could
extract spectra with good signal-to-noise ratio from two phases only:
the ``high'' phase spectrum by accumulating spectra near the crests, and
the ``low'' phase spectrum by accumulating spectra near the troughs. In
\S\ref{prepare_data} we describe in detail the algorithm we used to
create the phase-resolved ``high'' and ``low'' spectra.  The spectral
analysis is described in \S\ref{analysis}, and conclusions are summarized
in \S\ref{conclusion}.

\section{Data preparation} \label{prepare_data}
\begin{figure} 
\centering
\includegraphics[angle=-90, width=1.0\textwidth]{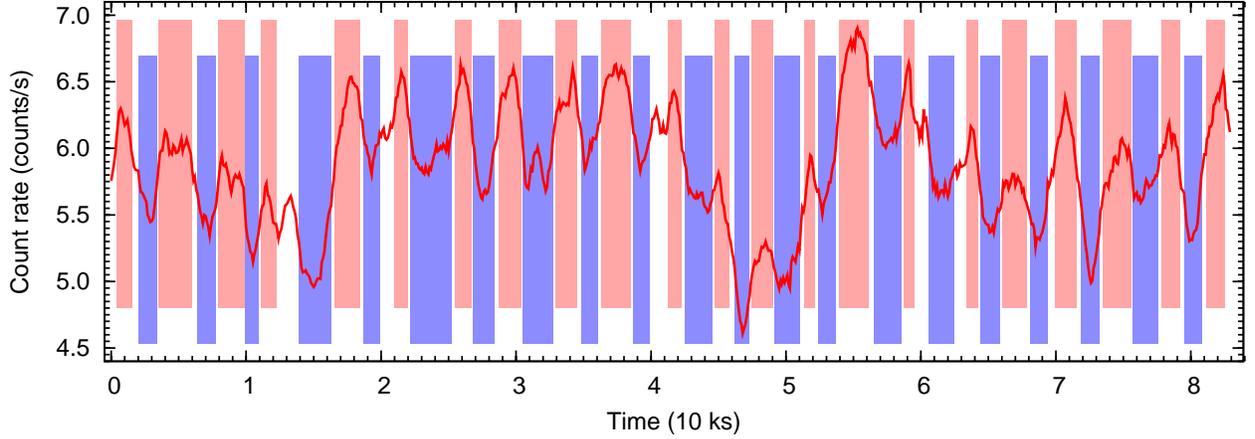}
\caption{
The red line shows the 9 point moving averaged light curve extracted
from EPIC PN data, from a circular region of 45 arcsecond radius around
the source, i.e. before excising the central, piled up core of the PSF.
The start time here corresponds to 2007 May 31, 20:11:01 UTC. The high and
low phases selected using the algorithm described in \S\ref{prepare_data}
are shown in light red and light blue respectively.  Please see the
electronic edition of the journal for a color version of this figure.
\label{fig:lc}}
\end{figure}

The QPO in \src\ was noticed in an \xmm\ observation starting on 2007
May 31 (observation ID. 0506440101).  The EPIC-PN detector was operated
in conjunction with the {\em thin} filter in {\em full-frame} mode
during this observation.  We extracted the light curves and spectral
information of this data set from the \epn\ detector using \xmm\ Science
Analysis Software (SAS; v.9.0.0) and following the \xmm\ {\em User Guide}
\footnote{http://xmm.esac.esa.int/external/xmm\_user\_support/documentation/sas\_usg/USG
}.  Briefly, we selected data from the PN detector which satisfies
the following criteria: pattern $\leq$4 (i.e. single and double pixel
events only), events with pulse height between 0.2 and 15 keV, FLAG=0
(the most stringent filter recommended to get a high-quality spectrum).
The red line in Fig.~\ref{fig:lc} shows the 9 point moving averaged
light curve extracted from EPIC PN data, from a circular region of 45
arcsecond radius around the source.  A circular, source-free region,
of radius 43.5 arcseconds on the same chip where the source was
located, was chosen to extract background spectra.  We used the SAS
task {\em epatplot} to anlyze the event pattern information near the
source region from the PN event file.  In the absence of any pile up,
the observed-to-model singles and doubles pattern fractions ratios
(obtained from {\em epatplot} output) should both be consistent with
1.0 within statistical errors.  Presence of pile up causes the singles
ratio to decrease from unity and the doubles ratio to increase from unity.
When a circular source region is chosen around the flux centroid of \src,
the observed distribution of single and double pattern fractions (as a
function of energy) is highly discrepant from the expected model curves.
The 0.5--2 keV observed-to-model fraction of singles in this case is
$0.905\pm0.003$ and that of the doubles is $1.298\pm0.007$, showing a high
pile up.  Therefore we excised the PSF core where pile up is strongest.
The optimal radius-of-exclusion was determined by slowly increasing the
exclusion radius till the observed distribution of single and double
events matched fairly well with the expected curves\footnote{See e.g.,
http://www.astro.lsa.umich.edu/$\sim$dmaitra/epat.gif which shows the
{\em epatplot} outputs for various exclusion radii.  The numerical
values of the radii are included in the file names shown on top right.
Please note that the radii are in units of 0.05 arcsec, i.e. XMM's
sky coordinates.}.  Thus we verified that the source region used by
\citet{middletonetal2009}, i.e. excluding the inner 32 arcsec and using
an annular region with outer radius of 45 arcsec, is an optimal choice.
The 0.5--2 keV observed-to-model fraction of singles after the exclusion
of inner 32 arcsec is $0.983\pm0.017$ and the corresponding fraction
of doubles is $1.096\pm0.028$.  The average PN source count-rates in
0.3--10 keV energy range before and after excluding the PSF core were
$5.8$ counts/s and $0.31$ counts/s respectively.  Appropriate photon
redistribution matrices and ancillary region files were created using
{\em rmfgen} and {\em arfgen} tasks.

\begin{figure} 
\includegraphics[angle=-90, scale=0.65]{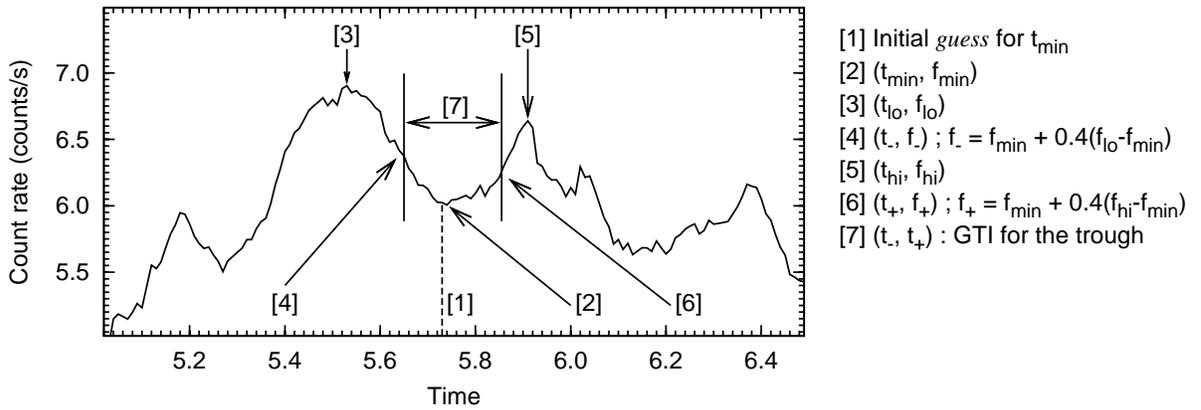}
\caption{An example showing the selection of good time interval 
near a trough, following the algorithm described in \S\ref{prepare_data}. 
\label{fig:gti}}
\end{figure}

As noted by \citet{g08} (also see Fig.~\ref{fig:lc} where the 9-point
moving average light curve computed from the 100-s binned PN data is shown
by the thick red line), a visual inspection of the light curve clearly
shows the periodicity of $\sim$1 hour.  These authors also noted that
the variability is not strictly periodic over the entire observation,
in particular during the initial $\sim$25 ks of the observation.
Therefore we adopted the following algorithm (also illustrated in
Fig.~\ref{fig:gti}) to extract spectral data from the ``high'' (crest)
and ``low'' (trough) phases.  Let the 9-point moving averaged flux at
time $t$ be denoted by $f(t)$.
\\(1) In order to select a trough in the flux, we start with an initial
approximation of the time of the trough, say $t_{\rm i}$.
\\(2) We then search the interval $\left( t_{\rm i}-P/4, t_{\rm i}+P/4
\right)$, $P$ being the QPO period, for a local minimum $f_{\rm min}$,
occurring at time $t_{\rm min}$.
\\(3) Next we search for the crest immediately preceding $ t_{\rm min}$,
occurring at $t_{\rm lo}$ with flux $f_{\rm lo}$.
\\(4) Now starting from $t_{\rm min}$, we search backwards in time
within the interval $ \left( t_{\rm lo}, t_{\rm min} \right)$ for the
time $t_{-}$ when the flux reaches $f_{-}\equiv f_{\rm min} + \eta \cdot
\left( f_{\rm lo} - f_{\rm min} \right)$, with $\eta=0.4$.  This gives
the left endpoint of the ``good time interval'' (GTI) for the trough at
$t_{\rm min}$.
\\(5) To find the right endpoint of the good time interval for the trough,
we similarly search for the crest immediately succeeding $t_{\rm min}$,
say $t_{\rm hi}$.
\\(6) As in the construction of $t_{-}$, we linearly interpolate starting
from $t_{min}$ and search forward to find the time $t_{+}$ when the
flux reaches $f_{+}\equiv f_{\rm min} + \eta \cdot \left( f_{\rm hi} -
f_{\rm min} \right)$.
\\(7) We now have a good time imterval $\left( t_{-}, t_{+} \right) $
for the trough of $t_{\rm min}$. A similar method is employed to estimate
the GTI for the crests.

\noindent At the boundaries of the light curve, we assumed that both the
endpoints were local minima and computed the GTI for their respective
neighboring local maxima.  Looking at the light curve in Fig.~\ref{fig:lc}
it is clear that this allows a conservative estimation of the GTIs near
the boundaries.

The advantages of using this algorithm are: (1) it allows better treatment
of asymmetric extrema, i.e. either when the maxima/minima are broad
(e.g. the crest near $t$=0.5) or when the two crests (troughs) that
enclose a given trough (crest) have unequal amplitudes (e.g.  the trough
near $t$=1.5); (2) it allows better treatment of light curves that are
not strictly periodic but slightly wander in phase or frequency.

The parameter $\eta$ essentially determines the width of each GTI around
any given crest or trough. The choice of its value of 0.4 in this case is
primarily dictated by photon statistics. A value smaller than 0.4 would
select narrower strips near the crests/troughs, and in principle should
give a sharper distinction between spectra from crests and troughs.
In reality however we are limited by photon number statistics and the
signal-to-noise of the spectra starts becoming poorer for smaller $\eta$.
Increasing $\eta$ on the other hand (e.g. $>$0.5) causes the strips to
overlap which is obviously not intended. Thus 0.4 is an optimal choice
for $\eta$.

The selected high and low phases are shown in light red and light blue
respectively in Fig.~\ref{fig:lc}. The algorithm chooses all the extrema
correctly except the ones near $t$=1.3, 4.1 and 6; it fails in these
cases because the successive extrema occur in time scales significantly
shorter than expected (nominally we expect two extrema to be separated
by P/2). We have excluded these three time intervals from our analysis.
The task {\em gtibuild} was used to create SAS-compatible GTI files for
both high and low phases, and these GTI selections were passed to {\em
evselect} task to extract the phase-resolved spectra.

\section{Analysis} \label{analysis}
We used the ISIS (v.1.5.0-7; see \citealt{isisref}) spectroscopy
package to analyze the spectra presented here.  In combination
with additional tasks written by Mike Nowak\footnote{Available at
http://space.mit.edu/home/mnowak/isis\_vs\_xspec/download.html},
ISIS offers flexible options\footnote{Such as rebinning (by minimum
counts/bin or minimum signal-to-noise ratio/bin) and setting systematic
errors on the fly, better handling of `unfolded spectra', its support
for parallel processing to take advantage of multi-cpu/cluster computing
environment allowing fast computation of numerically extensive tasks
such as estimation of confidence limits for model parameters.}, which
made it a good choice for the analysis presented here.
We grouped the \epn\ data for all fits starting at $\geq$0.25 keV,
with the criterion that the signal-to-noise ratio in each bin had to
be $\geq${\bf 5}. We then only considered data in the interval 0.25--10
keV. The photons statistics in the phase-resolved spectra, particularly
those of the higher energy channels, are low because of (a) the low
count rates after pileup correction, (b) keeping single and double
pixel events only and using FLAG=0 to ensure highest quality spectrum,
(c) the stringent S/N requirement in any bin, and obviously (d) the
phase grouping.
Galactic absorption was fixed to $N_H=1.31\times10^{20}$ cm$^{-2}$
in all fits.  The quoted errors correspond to 90\% confidence limit,
unless explicitly stated otherwise.

\subsection{Fits to the high phase spectrum}\label{s:hi} 
A high-resolution X-ray spectrum of the heart of an active galactic
nucleus, particularly a NLS1, has a complex shape, and physically
motivated models must include contributions from an accretion disk
\citep{ss1973,watarai2000}, Compton upscattering of soft photons from the
disk \citep{titarchuk1994}, reflection \citep{rossfabian1993}, disk winds
\citep{pk2004,sd2007}, and also sometimes partial covering of the source
by clumpy clouds near the source \citep[e.g.][]{reeves2008}.  Furthermore,
in comparison with most of the other active galactic nuclei \src\ shows a
very strong soft X-ray excess whose origin is yet to be fully understood
\citep[see e.g.,][for discussions]{middleton2007,middletonetal2009}.
Here however, our initial question is whether the continuum itself
changes shape between the high and low phases. Any continuum model
which fits one of the phases well can be used to answer this question.
Hence we restrict ourselves to simple continuum models.

\begin{figure} 
\centering
\plottwo{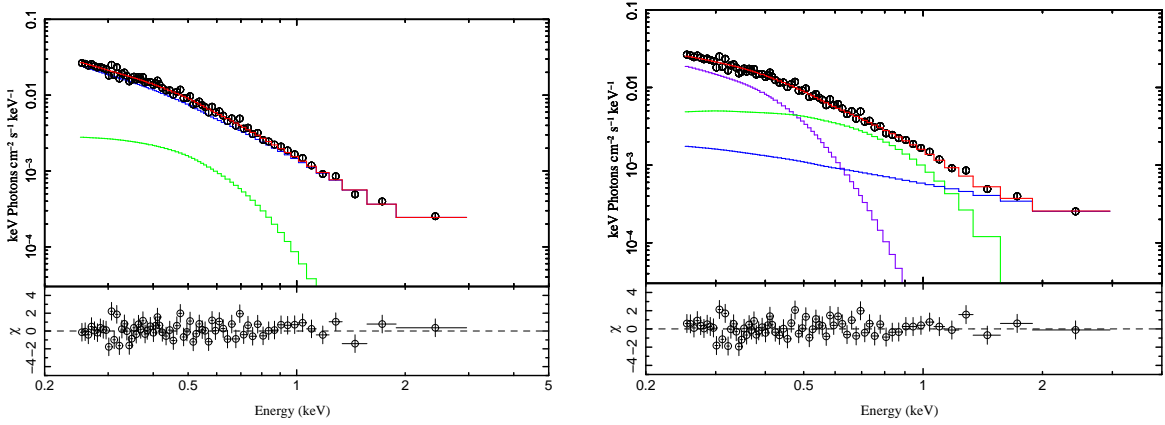}{hi_2zbb_pow.ps}
\caption{
{\em Left panel:} Best fit blackbody+broken power-law model
($Mhi\_bb\_bkn$ in Table~\ref{tab:fits}) and residuals to the high phase
spectrum of \src.  The power-law component is shown by the blue histogram,
the blackbody in green, and the total in red.
{\em Right panel:} Best fit two blackbodies and a power-law model
($Mhi\_2bb\_pow$ in Table~\ref{tab:fits}) and residuals to the same
dataset shown in the left panel.  The power-law component is shown by
the blue histogram, the high temperature blackbody in green, the low
temperature blackbody in violet, and the total in red. Please see the
electronic edition of the journal for a color version of this figure.
\label{fig:hi}}
\end{figure}

The high phase spectrum can be well modeled with a simple
blackbody of temperature $kT_{bb}= 0.1^{+0.03}_{-0.02}$ keV and
a broken power-law with photon index $\Gamma_1= 3.6\pm0.1 $ below
1.9 keV and $\Gamma_2=1.9_{-0.9}^{+0.4}$ above 1.9 keV (also see
Table~\ref{tab:fits}).  Hereafter this model will be referred to as
$Mhi\_bb\_bkn$.  While formally acceptable with \redc\ = 125.6/141, the
power-law component in this model dominates not only the high energies,
but also the low energies as shown in Fig.~\ref{fig:hi}.  Since only the
peak of the blackbody contributes to the observed spectrum, replacing
the blackbody with accretion disc models does not change the fits
significantly.  The fact that the power-law dominated both low and high
energies when using a single blackbody was also noted in a previous work
by \citet{pounds1995}.  Following \citet{pounds1995} we tried fitting two
blackbodies with different temperature and normalizations to fit the soft
($\sim$0.3--0.9 keV) and very soft ($<$0.3 keV) energies and a power-law
for the high ($>$0.9 keV) energies. The best fit parameters for this
model (hereafter $Mhi\_2bb\_pow$) are presented in Table~\ref{tab:fits},
and the model components and residuals are shown in Fig.~\ref{fig:hi}.
Note that the power-law model parameters $\Gamma_2$ for the model
$Mhi\_bb\_bkn$, and $\Gamma$ for the model $Mhi\_2bb\_pow$ are largely
constrained only by a few of the hardest energy bins, and consequently
have large error bars.  The 0.25--5 keV flux in the high phase spectrum
is $(1.26\pm0.1)\times10^{-11}$ ergs/cm$^2$/s.


\subsection{Fits to the low phase spectrum} \label{s:lo} 
First we test if the shape of the low phase spectrum is the same as that
of the high phase.  For this we multiply the best-fit high phase model by
an overall normalization constant and search for minimum $\chi^2$ allowing
only this constant to change.  The best-fit normalization constant has a
value in the range of $0.90\pm0.02$ for both models $Mhi\_bb\_bkn$ and
$Mhi\_2bb\_pow$, when applied to the low phase spectrum.  However the
residuals in this case (see Fig.~\ref{f:lo_renorm}) show a systematic
deviation in the $\sim$0.85--1.1 keV range, strongly suggestive of an
absorption edge.  There may also be some systematic deviation at lower
energies, especially an edge-like absorption feature between 0.3--0.4
keV. While not statistically significant in the current dataset, if
this indeed is the case then the continuum level at these energies
is higher than our estimate.  Some of these spectral features were
also noted by \citet{middletonetal2009} in the time-averaged spectrum,
but coadding spectra from both high and low phases most likely lowered
the observability of the features (which are prominent only during the
low phase).

Noting that the 0.8--0.9 keV feature could be due to an H-like oxygen
VIII K edge (at rest frame energy of 0.87 keV, seen in many Seyfert 1
sources \citep[see e.g.][]{reynolds1997} due to presence of ionized
gas in the line of sight), we next attempt to test if this feature
(present only during the low phase) could indeed be due to the oxygen
absorption edge.  For this we included a multiplicative \texttt{edge}
component to our models.  The best-fit values after including an edge in
both $Mhi\_bb\_bkn$ and $Mhi\_2bb\_pow$ are given in Table~\ref{tab:fits}.

The improvement in $\chi^2$ fit statistic for model $Mlo\_bb\_bkn\_edge$
between $\tau=0$ (no edge) and $\tau=0.29$ (best fit) is
$\Delta\chi^2=19.4$, implying a single parameter confidence limit of
$4.4\sigma$.
The improvement in $\chi^2$ fit statistic for $Mlo\_2bb\_pow\_edge$
between $\tau=0$ (no edge) and $\tau=0.23$ (best fit) is
$\Delta\chi^2=16.6$, implying a single parameter confidence limit of
$4.07\sigma$.

Thus both low phase models need an edge at $>$4$\sigma$ level, strongly
suggesting that the edge is present in the data.  While fits to both
models $Mlo\_bb\_bkn$ and $Mlo\_2bb\_pow$ are formally acceptable,
there remain hints of some systematic trend in the residuals shown in
Fig.~\ref{f:lo_renorm} at lower energies.  This is discussed in greater
detail in \S\ref{conclusion}.  The 0.25--5 keV flux in the low phase
spectrum is $(1.14\pm0.1)\times10^{-11}$ ergs/cm$^2$/s.

Next we attempted to estimate the strength of the same OVIII absorption
edge in the high phase data, by adding an edge to both the models
$Mhi\_bb\_bkn$ and $Mhi\_2bb\_pow$ and redoing the fits.  We found
that inclusion of the edge did not improve the statistical quality of
the fits. The OVIII edge could not be detected (at $>3\sigma$ level)
in the data using either of the continuum models, and this confirms that
the optical depth of the edge varied significantly between the high and
low phases.

Next we modeled the edge in the low phase somewhat more physically using
the XSTAR photoionization code \citep{xstar}. A grid of XSTAR models
was created from the following variables:
(1) the column density ($N_H$, allowed range $10^{20}$ -- $10^{23}$ cm$^{-2}$), 
(2) ionization parameter ($log(\xi)=log(L/(nR^2))$, allowed range 2--4),
(3) oxygen abundance (relative to solar, allowed range 0.1--4),
(4) neon abundance (relative to solar, allowed range 0.1--1). 
A multiplicative table model was created from this multidimensional grid.
Turbulent velocities of 300 km/s and above cause absorption lines to
become more prominent, and lower turbulent velocities are required
to fit the data well.  The best-fit parameters to the low phase data
using this model in conjunction with a continuum model composed of
double-blackbody+power law suggest $N_H$/(10$^{21}$ cm$^{-2}$) = $4\pm1$,
log($\xi$)=$ 3.2 _{-0.1}^{+0.2}$, and a relatively high oxygen abundance
($ 2.1 _{-0.5}^{+0.8} $). A sub-solar neon abundance (0.13) is preferred
by the best-fit solution but its error-bar could not be constrained. See
Table~\ref{tab:fits} for the detailed fit parameters and statistics. The
best-fit continuum parameters in this case were very similar to those
obtained using the single edge model ($Mlo\_2bb\_pow$).

It is possible that oscillations in the flux of the source cause the
ionization of intervening gas to vary, and that the gas is too ionized to
give significant OVIII absorption during high phases.  This possibility
can be tested using our XSTAR model for the warm absorber gas.  If this
were the case, then it ought to be possible to model the high phase
spectrum using the same warm absorber abundances as obtained for the
best-fit to the low phase data, but with a higher ionization fraction
($\sim$10--20\% higher, which is typically the flux difference between
the crests and peaks). However the high phase spectrum cannot be modeled
with 10--20\% enhanced ionization fraction.  In fact, detailed modeling
shows that reasonable fits to the high phase spectrum can be obtained
only if the ionization fraction is a factor of three higher than that
required by the low phase spectrum.  Thus it seems unlikely that the
changes in the high and low phase spectrum are driven solely by flux
viariations in the source.

We have also repeated the entire analyses using a different, somewhat
larger exclusion radius of 35 arcseconds (smaller exclusion radii were
not considered as they would be affected by pile up). In this case the
errors are larger due to poorer photon statistics, but the conclusions are
same, i.e. an edge is detected (at $>3\sigma$) level in the low phase,
but no edge is detected in the high phase.  Similarly we repeated the
analyses extracting only pattern=0 events (single pixel events), and
excising the PSF core of radius 30 arcseconds. In this case also the
OVIII edge was detected at 4.1$\sigma$ confidence level in the low phase
spectrum whereas no evidence of the OVIII edge could be found the high
phase spectrum.  Thus our main observation, viz. that the OVIII absorption
edge is {\em only} seen during the troughs in the oscillations, and not
during the crests, appears to be quite robust and is not some form of
detector artifact.



\section{Discussion and Conclusions} \label{conclusion}
In this work we have presented phase-resolved spectroscopy of the narrow
line Seyfert 1 galaxy \src. We have used the \xmm\ observation of this
source between 2007 May 31 and 2007 June 1, when the X-ray light curve
showed strong quasi-periodic oscillations.  We have extracted spectra
during the high (crest) and low (trough) phases.  Simple continuum
models assuming either a blackbody plus broken power-law, or two
blackbodies plus a power-law model, fit the high phase spectrum
quite well.  The required low-energy power-law index is very steep
($\Gamma\sim3.6\pm0.1$) for a single blackbody plus broken power-law
model, which is typical for narrow line AGNs with a high soft X-ray excess
\citep{boller1996,pounds1995,middleton2007}.  The spectrum can also be
well modeled using two blackbodies and a power-law continuum.  Due to
the steep decline of the spectrum, and high pileup on the PN detector
which severely limits the photon statistics, we do not have good quality
spectral data beyond $\sim$2--3 keV.  Therefore for these spectra we
cannot very well constrain the relatively flatter 2--10 keV continuum
usually seen in this source \citep{crummy2006,middletonetal2009},
although fits to a two-blackbody plus power-law model show that the $>2$
keV spectrum is comparatively flatter with a photon index of $\sim2$.

\begin{figure} 
\centering
\plottwo{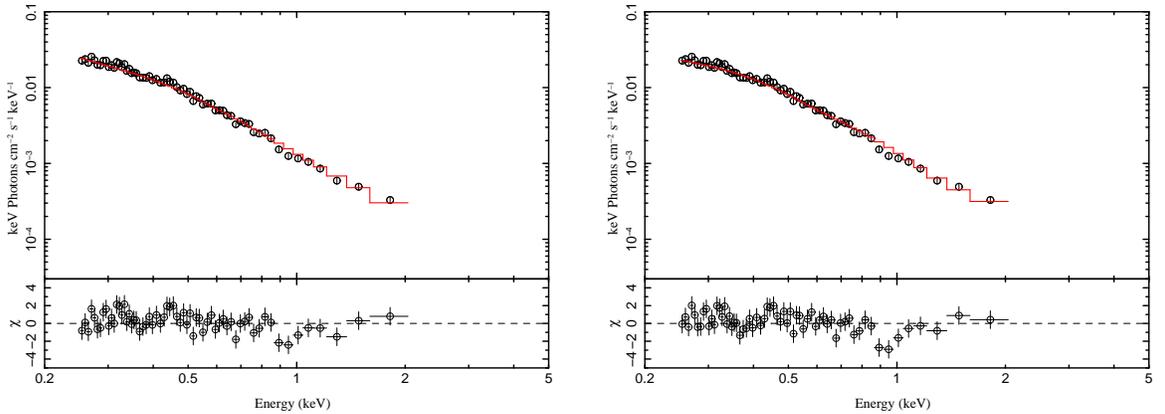}{lo_2zbb_pow_renorm.ps}
\caption{
{\em Left panel:} Best fit to the low phase data using the model
$Mhi\_bb\_bkn$ multiplied by an overall normalization constant and
allowing only the normalization constant to vary, thus keeping the
spectral shape same as $Mhi\_bb\_bkn$.
{\em Right panel:} Same as left panel but for model $Mhi\_2bb\_pow$.  
In both cases note the systematic deviations in the residuals in the
$\sim$0.85--1.1 keV range.
\label{f:lo_renorm}}
\end{figure}

\begin{figure} 
\centering
\plottwo{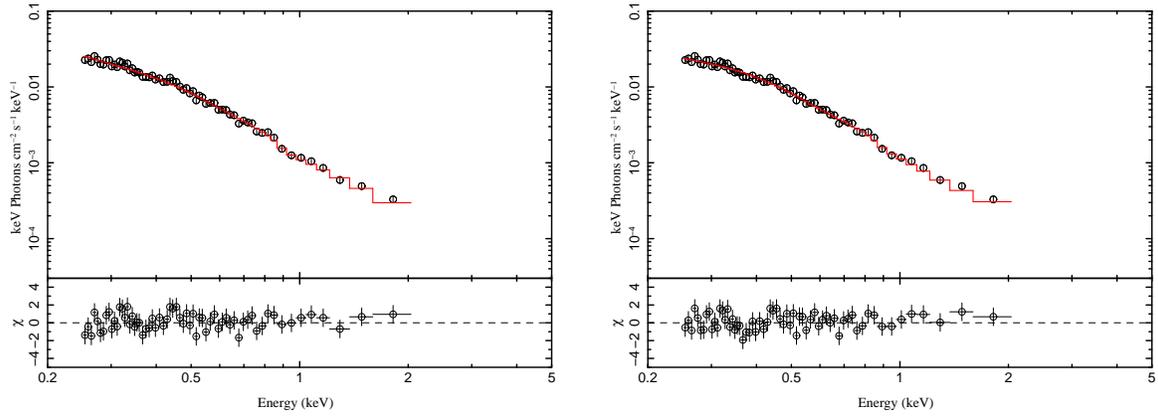}{lo_2zbb_pow_edge.ps}
\caption{Best fit to the low phase data using models $Mlo\_bb\_bkn\_edge$
(left panel) and $Mlo\_2bb\_pow\_edge$ (right panel) which includes
the (H-like) oxygen VIII K edge.  The best fit parameters are given
in Table~\ref{tab:fits}.
\label{fig:lo_edge}}
\end{figure}

The low phase spectrum cannot be adequately described by scaling the
overall normalization of either of the best fit high phase models,
suggeting that the shape of the low phase spectrum is different from that
of the high phase.  Analyzing the r.m.s. variability at different energies
\citet{middletonetal2009} also concluded that the variability is mainly
from the hard photons.  The right panel of Fig.~\ref{f:lo_renorm} shows
that the differences between the low and high phases are most prominent at
higher energies.  In particular, the residuals in Fig.~\ref{f:lo_renorm}
show a sharp drop in the continuum flux between 0.8--0.9 keV, which
then slowly increases toward the expected continuum flux at higher
energies, reminiscent of an absorption edge.  The inclusion of an
edge improves the fits significantly (at $>$4$\sigma$ level) for both
single-blackbody+broken power-law model and double-blackbody+power-law
model.  The best fit values for the edge threshold energy and absorption
depth are $0.87\pm0.04$ keV and $0.29\pm0.11$ respectively for the
single-blackbody+broken power-law model, and $0.86^{+0.05}_{-0.06}$ keV
and $0.23^{+0.12}_{-0.11}$ respectively for the double-blackbody+power-law
model.  The edge is most likely associated to the 0.87 keV (rest frame)
K-edge of (H-like) oxygen VIII which is quite commonly seen in Seyferts
\citep[see e.g.,][]{reynolds1997}, and is indicative of the presence
of optically thin, photoionized matter along the line of sight.  \src\
however is at a redshift of 0.042; therefore the 0.87 keV edge (in rest
frame) should appear at 0.84 keV (easily within the 90\% confidence error
bar for both models). The presence of a warm absorber should create other
features in the spectrum, e.g. the 0.74 keV K edge from He-like O VII.
While the residuals in Fig.~\ref{fig:lo_edge} do show a drop in the flux,
it is not very significant statistically.

To estimate the column density ($N_{\rm OVIII}$) of the O VIII ions
responsible for the edge we used Eq.(1) of \citet{verneretal1996} and
calculated the absorption cross section $\sigma$ (for O VIII, $\sigma[{\rm
E}=0.87\;{\rm keV}]=0.098$ Mb).  Since $\tau=N_{\rm OVIII}\cdot\sigma$,
we obtain $N_{\rm OVIII}=3\times10^{18}$ cm$^{-2}$.  Estimating the
number density ($n_{\rm OVIII}$) requires a size-scale for the warm
absorber.  Assuming that the QPO originates in the accretion disk, the
warm absorber is could be confined within the orbit of the QPO, or in
case of an outflow it originates inside the QPO orbit. The mass of the
supermassive black hole in \src\ is in the range of $1-7\times10^{6}$
\msun\ \citep{BianHuang2010,zhouetal2010}.  A circular Keplerian
orbit would have a radius of $9.4(M/[4\times10^6 \msun])^{-2/3}(P/[1
hr])^{2/3}\; r_{\rm g}$, which would be an upper limit on the size of
the absorber in this case. This would give $n_{\rm OVIII}=3\times10^5$
cm$^{-3}$ for the number density of OVIII ions.  Note however that this
assumes a Keplerian orbit origin for the QPO, and therefore depends on
the mass of the central supermassive black hole.  Thus in \src\ we may
be seeing an infalling ``blob'', the emission from which is periodically
absorbed as it passes behind a warm absorber in the immediate vicinity
of the central engine.  If the ``blob'' is at $\sim$9 $r_{\rm g}$, that
would imply the ionizing flux must be generated within 9 $r_{\rm g}$
(or else we would not see absorption), and this would be easier to
accomplish if the black hole is spinning since the disk can get down
to a smaller radius (e.g. as low as 1.25 $r_{\rm g}$ for a maximally
spinning black hole). Recent work by \citet{crummy2006} has shown that
NLS1 spectra can be modeled using reflection from a disk around a spinning
black holes \citep[see][for a review]{Miller2007}.

We also created a multidimensional grid of models using the XSTAR
photoionization code, to create a more physical model for the
warm absorber and explore a wide range of column density, ionization
parameter, oxygen and neon abundance. Fits to the low-phase data using
this model suggests a warm absorber column density of $\sim4\times10^{21}$
cm$^{-2}$, log($\xi$)=$3.2 _{-0.1}^{+0.2}$, and an oxygen abundance of
$2.1 _{-0.5}^{+0.8}$ relative to solar.  The neon abundance could not be
well constrained from the data, although the best-fit solution suggests
a sub-solar value.

Another possible physical scenario for the difference between high and
low phase spectrum could be that the warm absorber is along the line
of sight, between us and the source (where the QPO originates), and the
warm absorber responds to flux variations in the source itself, i.e. it
is more ionized and transparent at high fluxes and less at lower fluxes.
However, based on our modeling of the spectra using physically motivated
models generated using XSTAR this scenario appears unlikely because the
change in luminosity of the ionizing flux between low and high phases is
too small to create the observed difference in the depth of the OVIII edge
(also see \S\ref{analysis}).

It is interesting to note in this context that a somewhat complementary
situation has been observed in the Seyfert galaxy NGC~1365, where
\citet{Risalitietal2009} have reported a possible transit of an obscuring
cloud (with $N_H\sim3.5\times10^{23}$ cm$^{-2}$ and other inferred
properties similar to that of a broad-line region cloud) in front of
the central X-ray source.

Signatures of variable absorption on short timescales have been previously
observed in high resolution {\em Chandra} spectra of the stellar black
hole candidate H1743--22 \citep{miller2006}.  For H1743--22, the lines
were slightly blueshifted, suggesting an outflow origin. For \src, we
do not have definitive evidence for an outflow, but the best fit edge
energy for both models is higher (though within 90\% confidence interval)
than the rest frame energy of the oxygen edge.  If the warm absorber
in \src\ is indeed in an outflow, then the observed oscillations could
originate due to an instability in the inner accretion disk threaded
by a poloidal magnetic field, giving rise to an outflowing wind or jet
\citep{BlandfordPayne1982,tagger1999}.  The observed oscillations in
this case would correspond to scaled up versions of low-frequency QPOs
\citep[see e.g.,][]{vdk2006} seen in stellar mass X-ray binaries.

Since the presence of a warm absorber is bound to have noticeable
imprints on other regions of the electromagnetic spectrum, especially
in UV and soft X-rays, future deep UV and X-ray observations of
\src\ will be useful not only in constraining the origin of the
oscillations, but also in gaining a better understanding of the
physical emission mechanism of sources like \src\ which show a strong
soft-X-ray excess. Future X-ray missions with improved sensitivity and
larger collecting area like the {\em International X-ray Observatory}
\citep[IXO;][]{Milleretal2009,Whiteetal2010} would play key role in
understanding accretion geometry as well as emission/absorption mechanism
in these sources.


\acknowledgments
DM thanks Mike Nowak for making his analysis scripts publicly available,
and Ryan Porter and Eric Pellegrini for lively discussions. We would like
to thank the anonymous referee for valuable suggestions which have helped
to improve the paper. This research has made use of data obtained from
the High Energy Astrophysics Science Archive Research Center (HEASARC),
provided by NASA's Goddard Space Flight Center. This work is based on
observations obtained with XMM-Newton, an ESA science mission with
instruments and contributions directly funded by ESA Member States
and NASA.

{\it Facilities:}  \facility{XMM-Newton}.


\begin{deluxetable}{lcll}
\tabletypesize{\small}
\tablecaption{Summary of fits.  
\label{tab:fits}
}
\tablewidth{0pt}
\tablehead{
\colhead{Phase} & \colhead{Model\tablenotemark{1}} & 
\colhead{Best-fit parameters\tablenotemark{2}} & \colhead{Best-fit statistics}
}
\startdata
High	& $Mhi\_bb\_bkn$  
        & $kT_{bb} = 0.1 _{-0.02}^{+0.03} $ & \\
 &	& $N_{bb}/10^{-5} = 3.2 _{-1.7}^{+2.9} $ & \\
 &	& $\Gamma_1 = 3.6\pm0.1 $ & \\
 &	& $N_{p}/10^{-3} = 1.5\pm0.1 $ & \\
 &	& $E_b = 1.9 _{-0.3}^{+0.6} $ & \\
 &	& $\Gamma_2 = 1.9 _{-0.9}^{+0.4} $ & \\
 & & & $\chi^2/\nu=125.6/141$\\
\hline
High	& $Mhi\_2bb\_pow$ 
        & $kT_{bb1} =0.06\pm0.01 $ & \\
 &	& $N_{bb1}/10^{-4} = 2.8^{+0.4}_{-0.3} $ & \\
 &	& $kT_{bb2} =0.14\pm0.02 $ & \\
 &	& $N_{bb2}/10^{-5} = 6.9^{+1.6}_{-1.3} $ & \\
 &	& $\Gamma = 2.0\pm0.4 $ & \\
 &	& $N_{p}/10^{-4} = 6^{+3}_{-2} $ & \\
 & & & $\chi^2/\nu=130.0/141$ \\
\hline
Low	& $Mlo\_bb\_bkn\_edge$	
        & $ kT_{bb} = 0.09 _{-0.01}^{+0.01} $ & \\
 &	& $N_{bb}/10^{-5} = 2.9 _{-0.4}^{+0.5} $ & \\
 &	& $N_{p}/10^{-3} = 1.3\pm0.2 $ & \\
 &	& $\Gamma_1 = 3.6\pm0.03 $ & \\
 &	& $E_b = 1.8 _{-0.2}^{+0.7} $ & \\
 &	& $\Gamma_2 = 2.3\pm0.5 $ & \\
 &	& $E_e$ = $0.87\pm0.04$ & \\
 &	& $\tau = 0.29\pm0.11 $ & \\
 & & & $\chi^2/\nu=136.0/128$ \\
\hline
Low	& $Mlo\_2bb\_pow\_edge$ 
        & $kT_{bb1} = 0.06 _{-0.02}^{+0.01} $ & \\
 &	& $N_{bb1}/10^{-4} = 2.4_{-0.0}^{+0.1} $ & \\
 &	& $kT_{bb2} = 0.14\pm0.04 $ & \\
 &	& $N_{bb2}/10^{-5} = 5.6\pm0.2 $ & \\
 &	& $\Gamma = 2.36 _{-0.58}^{+0.06} $ & \\
 &	& $N_{p}/10^{-4} = 6.7_{-2.7}^{+0.4} $ & \\
 &	& $E_e$ = $ 0.86 _{-0.06}^{+0.05} $  & \\
 &	& $\tau = 0.23 _{-0.11}^{+0.12}  $ & \\
 & & & $\chi^2/\nu= 133.6/128$ \\
\hline
Low	& $Mlo\_2bb\_pow\_xsgrid$ 
        & $kT_{bb1} = 0.06\pm0.01$ & \\
 &	& $N_{bb1}/10^{-4} = 2.4_{-0.0}^{+0.1} $ & \\
 &	& $kT_{bb2} = 0.14\pm0.01 $ & \\
 &	& $N_{bb2}/10^{-5} = 5.6_{0}^{+0.2} $ & \\
 &	& $\Gamma = 2.42 _{-0.51}^{+0.01} $ & \\
 &	& $N_{p}/10^{-4} = 7.4\pm0.4 $ & \\
 &	& $N_{H}/10^{21} = 4\pm1 $ & \\
 &	& log($\xi$) = $3.2 _{-0.1}^{+0.2}$ & \\
 &	& O$_{\rm Abund} = 2.1 _{-0.5}^{+0.8}$ & \\
 &	& Ne$_{\rm Abund} = 0.14 _{-0.04}^{+0.86}$ & \\
 & & & $\chi^2/\nu= 134.3/126$\\
\hline
\enddata
\tablenotetext{1}{ In ISIS/XSPEC notation the model definitions are:
$Mhi\_bb\_bkn$ $\equiv$ \texttt {phabs (zbbody + bknpower)}; 
$Mhi\_2bb\_pow$ $\equiv$ \texttt{phabs (zbbody(1) + zbbody(2) 
+ powerlaw)};
$Mlo\_bb\_bkn\_edge$ $\equiv$ \texttt{constant [phabs (edge (zbbody 
+ bknpower))]};
$Mlo\_2bb\_pow\_edge$ $\equiv$ \texttt{constant [phabs (edge (zbbody(1) 
+ zbbody(2) + powerlaw))]};
$Mlo\_2bb\_pow\_xsgrid$ $\equiv$ \texttt{constant [phabs (xsgrid (zbbody(1) 
+ zbbody(2) + powerlaw))]}
}
\tablenotetext{2}{
The parameter $kT_{bb}$ is the blackbody temperature
in keV.
The blackbody normalization $N_{bb}=L_{39}/[D_{10}(1+z)]^2$, where
$L_{39}$ is the source luminosity in units of $10^{39}$ erg/s,
$z=0.042$ is the redshift, and $D_{10}$ is the distance to the
source in units of 10 kpc.
$N_{p}$ is the number of power-law photons/cm$^2$/s/keV at 1 keV, 
and $\Gamma$ is the photon index of the power-law. 
When using a broken power law model 
$\Gamma_1$ is the photon index for $E<E_b$, and 
$\Gamma_2$ is the photon index for $E>E_b$, where $E_b$ is the 
break energy in keV.
The absorption edge is parametrized by its threshold energy ($E_e$, 
in keV) and absorption depth at the threshold ($\tau$).
The model \texttt{xsgrid} is a multiplicative table model created from and
XSTAR grid with following variables: column density ($N_H$), ionization
parameter (log($\xi$)), oxygen abundance relative to solar (O$_{\rm
Abund}$), and neon abumdance relative to solar (Ne$_{\rm Abund}$).
See \S\ref{analysis} for details of the model.
Galactic absorption was fixed to $N_H=1.31\times10^{20}$ cm$^{-2}$ in all
fits. The errors correspond to 90\% confidence limit.
}
\end{deluxetable}


\end{document}